\documentclass[aps,twocolumn,prd,showpacs,nofootinbib]{revtex4}
\usepackage{amsmath}
\usepackage{graphicx}
\usepackage{dcolumn}
\usepackage{bm}
\usepackage{amssymb}
\usepackage{latexsym}

\def\be{\begin{equation}}
\def\ee{\end{equation}}
\def\ba{\begin{eqnarray}}
\def\ea{\end{eqnarray}}

\begin{document}

\title{Primordial Perturbation in Horava-Lifshitz Cosmology
}

\author{Yun-Song Piao$^{a,b}$}
\affiliation{${}^a$College of Physical Sciences, Graduate School
of Chinese Academy of Sciences, Beijing 100049, China}
\affiliation{${}^b$Kavli Institute for Theoretical Physics China,
ITP-CAS, Beijing, P.R. China}

\begin{abstract}

Recently, Ho\v{r}ava has proposed a renormalizable theory of
gravity with critical exponent $z=3$ in the UV. This proposal
might imply the scale invariant primordial perturbation can be
generated in any expansion of early universe with $a\sim t^n$ and
$n>1/3$, which, in this note, will be validated by solving the
motion equation of perturbation mode on super sound horizon scale
for any background evolution of early universe.
However, it is found that it seems that if we require the efolding
number of primordial perturbation is suitable for observable
universe, $n\gtrsim 1$ still need to be satisfied, unless the
scale of UV regime is quite low.


\end{abstract}

 \maketitle

Recently, Ho\v{r}ava has proposed a renormalizable theory of
gravity at a Lifshitz point \cite{H}, which can be a UV complete
candidate of general relativity. In the UV, this theory has the
critical exponent $z=3$, at which the space and time scale
differently, which describes the interacting nonrelativistic
gravitons at short distances and is renormalizable. In the IR,
this theory flows naturally to the relativistic value $z=1$, and
the general relativity is recovered.

Recently, the Ho\v{r}ava-Lifshitz (HL) gravitation theory has been
studied intensively in Refs.
\cite{V},\cite{TS},\cite{C},\cite{KK},\cite{LMP},\cite{M3},
 \cite{B},\cite{N1},\cite{N},\cite{CCO}.
In Refs.\cite{LMP},\cite{CCO}, the black hole solutions were
studied. In Refs.\cite{C},\cite{KK},\cite{LMP}, the cosmological
solutions was explored. It was found that the early universe in HL
cosmology may be able to escape singularities and has a
nonsingular bounce. This might give an alternative to inflation,
as has been discussed in \cite{B} based on matter bounce. In Ref.
\cite{M3}, it was pointed out that, in UV regime of HL gravity,
the spectrum of primordial perturbation induced by a scalar field
may be scale invariant for any expansion with $a\sim t^n$ and
$n>1/3$.
This result is interesting. However, it might be required, and
also significant to show it by solving the motion equation of
perturbation mode on super sound horizon scale for any background
evolution of early universe, which in some sense helps to
understand how the perturbation generated in UV regime is matched
to the observations on large IR scale. This will be done in this
note.

In the UV of HL gravity, the action of a scalar mode, e.g. $\Phi$,
should have the critical exponent $z=3$, which likes, see
Refs.\cite{C},\cite{KK},\cite{M3} for more discussions, \be
{I}_{UV}\sim \int dtd{\rm
x}^3\left({\dot\Phi}^2+{\Phi\bigtriangleup^3\Phi\over
a^6M^4}\right), \label{I}\ee where $\bigtriangleup\equiv
\partial_i\partial_i$ is the spacial Laplacian, $a$ is the scale factor
and $M$ is the mass scale. The sign before $
{\Phi\bigtriangleup^3\Phi\over a^6M^4}$ is positive, which is
required by the stability in the UV. In general, the term $
{\Phi\bigtriangleup^3\Phi\over a^6M^4}$ is important only when
$k/a\gtrsim M$, which means the physical wavelengths of the
perturbation mode is quite short. This is consistent with the case
of a sufficiently early period of expanding universe. When $k/a\ll
M$, which occurs after the universe expands some time, the term $
{\Phi\bigtriangleup^3\Phi\over a^6M^4}$ will be replaced with
$\Phi\bigtriangleup\Phi$, which means the field theory flows to
the relativistic value $z=1$, where the space and time will scale
samely and the usual relativistic field theory will be acquired.
The motion equation of perturbation of $\Phi$ in the UV regime is
given by, in the momentum space, \be u_k^{\prime\prime}
+\left(\omega^2-{a^{\prime\prime}\over a}\right) u_k = 0
,\label{uk1}\ee where $u_k$ is related to the perturbation
$\delta\Phi$ of $\Phi$ by $u_k \equiv a\delta\Phi_k$ and the prime
denotes the derivative with respect to the conformal time $\eta$,
and $\omega={k^3\over a^2M^2}$. In principle, we may generally
take $\omega= {k^{z}\over (Ma)^{z-1}} $ for the calculation of
Eq.(\ref{uk1}), by which the perturbation spectrum can be obtained
for different values of $z$, which will be used in Eqs.(\ref{uk})
and (\ref{v}). We, no loosing generality, will take $a\sim t^n$
for calculations, where $n$ is a positive constant, thus for the
conformal time $a\sim \eta^{n\over 1-n}$.


The emergence of primordial perturbation in HL cosmology can be
explained as follows. The universe is initially in the UV regime
of HL gravity, $\omega={k^3\over a^2M^2}$. In this regime, since
$a$ is quite small, $\omega\eta \gg 1 $, the perturbation modes
can be regarded as adiabatic. The reason is that since
$\omega^{\prime}/\omega \sim 1/\eta$, thus the adiabatic condition
$\omega^{\prime}/\omega^2\ll 1$ is equivalent to $\omega\eta \gg 1
$. Noting $\eta\sim 1/(ah)$, we have $\omega\eta\sim
\omega/(ah)\gg 1$, and thus obtain $a/\omega \ll 1/h$, which
corresponds to the case that the effective physical wavelength is
quite deep into the horizon. Thus in this case, i.e. $\omega\eta
\gg 1 $, \be u_k\simeq {1\over
\sqrt{2\omega(k,\eta)}}\exp{(-i\int^{\eta}\omega(k,\eta)d\eta)}
\label{in}\ee can be regarded as an approximate solution of Eq.
(\ref{uk1}).
$\omega\eta $ will decrease with the expansion of $a$. Thus at
late time, we can expect $\omega\eta \ll 1$, i.e. $a/\omega\gg
1/h$, which means that the effective wavelength will evolve faster
than that of $1/h$, and thus will leave the horizon after some
time, see the red dashed lines in Fig.1. This condition that
$\omega\eta$ decreases with the time equals that $a^3h$ increase
with the time, since $\omega\sim 1/a^2$ and $\eta\sim 1/(ah)$.
Thus when considering $a\sim t^n$, $n>1/3$ is required \cite{M3},
which corresponds to the state equation $w<1$. While it is clear
that if the universe is in contraction, which corresponds to
$a\sim (-t)^n$ and in which the following Eqs.(\ref{uk}) and
(\ref{v}) can be also applied, the condition that the perturbation
is able to leave the horizon is $n<1/3$, i.e. $w>1$.

\begin{figure}[t]
\begin{center}
\includegraphics[width=8cm]{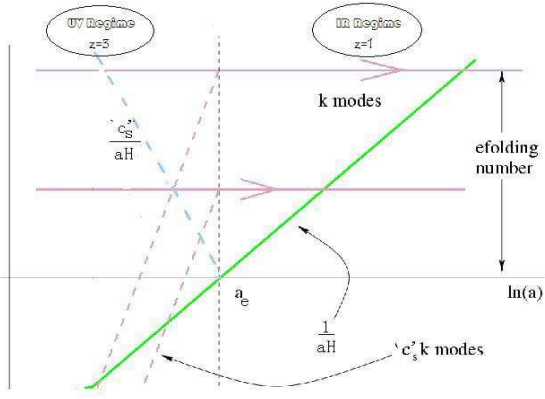}
\caption{ The figure of $\ln{({1\over ah})}$ with respect to
$\ln{a}$, see the green solid lines. The red solid lines are the
perturbation modes with wave number $k$. The ``effective" sound
speed $c_s= {k^2\over a^2M^2}$. The blue dashed line is that of
$\ln{({c_s\over ah})}$. The left side of $a_e$ is the UV regime in
which $z=3$ and its right side is the IR regime in which $z=1$.
$a_e$ denotes the time when the UV regime ends. In the UV regime,
initially $\omega\eta \gg 1 $, we have $a/k\ll c_s/h$. Thus though
$a/k\gg 1/h$, i.e. the physical wave length of perturbation mode
is larger than the horizon, it is actually smaller than the
``effective" sound horizon $c_s/h$, since ${k^2\over a^2M^2}$ is
large. Thus a causal relation can be established on super horizon
scale. When the universe expands, ${k^2\over a^2M^2}\sim 1/a^2$ is
decreased, thus the corresponding mode will leave this
``effective" sound horizon and can be able to be responsible for
seeds on large IR scale. }
\end{center}
\end{figure}

This result can be also explained in another perspective, see
Fig.1. When $\omega\eta \gg 1 $, we have $a/k\ll ({k^2\over
a^2M^2})/h$. Thus though $a/k\gg 1/h$, i.e. the physical wave
length of perturbation mode is larger than the horizon, it is
actually smaller than the ``effective" sound horizon $c_s/h$,
where the effective sound speed is defined as $c_s={k^2\over
a^2M^2}$ since ${k^2\over a^2M^2}$ is large. Thus a causal
relation can be established on superhorizon scale. When the
universe expands, ${k^2\over a^2M^2}\sim 1/a^2$ is decreased, thus
the corresponding mode will leave the ``effective" sound horizon
and can be able to be responsible for seeds in observable
universe. The decrease of sound horizon $c_s/h$ with the time
requires $a^3h$ is increased with the time, thus $n>1/3$ is
obtained. In this sense, the generation of primordial perturbation
in HL cosmology is similar to that in the scenario with the
decaying speed of sound \cite{Picon},\cite{Piao0609},\cite{M}, see
also Ref. \cite{Piao0807} for further illustration.

Eq.(\ref{uk1}) is a deformed Bessel equation, which is slightly
different from that used in usual calculations for perturbation,
since here ${k^2\over a^2M^2}$ before $k^2$ is rapidly changed
with the time. However, for primordial perturbation, such equation
has been solved in Ref. \cite{Piao0609} in detail. The general
solutions of Eq.(\ref{uk1}), which can be matched to Eq.(\ref{in})
when the mode $u_k$ is quite deep inside the sound horizon, i.e.
$\omega\eta\gg 1$, are the Hankel functions with the order $v$ and
the variable $\omega\eta$. This solution on super sound horizon
scale, i.e. $\omega\eta\ll 1$, is \ba u_k &\simeq &
{1\over \sqrt{2\omega}}\left(\omega\eta\right)^{0.5-v}\nonumber\\
& \simeq & {aM\over \sqrt{2k^3}}\left({k^2\over
a^2M^2}k\eta\right)^{0.5-v}, \label{uk}\ea which corresponds to
the expansion of Hankel functions to the leading term of
$\omega\eta$, where the prefactor of order one has been neglected,
the upper equation is that for general $\omega$, while the lower
equation is that for $z=3$. In principle $v$ in this deformed
Bessel equation is determined not only by $a^{\prime\prime}/a$,
but also by the dependence of $\omega$ on time. $v$ has been
calculated in Ref. \cite{Piao0609}, which is \be
v=0.5\left|{3n-1\over nz-1}\right| \label{v}\ee for any $z$. It
can be noticed that if $z=1$, $v$ will be reduced to the usual
result, in which only when $n\gg 1$ or $n={2/3}$, which correspond
to that of the inflation and the contraction dominated by matter
\cite{Wands99},\cite{FB}, see also earlier \cite{S}, respectively,
$v=1.5$ and thus ${\cal P}^{1/2}_{\Phi}\simeq
k^{3/2}\left|{u_k\over a}\right|$ is scale invariant, which is
familiar result. While if $z=2$, which corresponds to
$\omega={k^2\over aM}$, in term of Eq.(\ref{v}) and considering
${\cal P}^{1/2}_{\varphi}\simeq k^{3/2}\left|{u_k\over a}\right|$,
it can be found that the scale invariance of spectrum requires
$n\gg 1$ or $n={5/12}$. $n\gg 1$ apparently corresponds to that of
inflation. When $z=2$, for an expanding universe, the condition
that the perturbation is able to leave the horizon requires that
$a^2h$ increases with the time. This means $n>1/2$. Thus the case
with $n={5/12}$ actually corresponds to the contraction dominated
by the component with $w=3/5$. In principle, in term of
Eqs.(\ref{uk}) and (\ref{v}), we can deduce in what background
evolution of early universe, i.e. what value $n$ is, the
perturbation spectrum generated is scale invariant for some
special value of $z$ \footnote{The spectral index is actually
$n_s-1=3-z\left|{3n-1\over nz-1}\right|$. }.

Here $z=3$, thus it is found that $v\equiv 0.5$ and thus $u_k\sim
1/k^{3/2}$ for any value of $n$. The spectrum of primordial
perturbation induced by $\varphi$ is given by \be {\cal
P}^{1/2}_{\Phi}\simeq k^{3/2}\left|{u_k\over a}\right|\simeq M.
\label{pk}\ee Thus on super sound horizon scale, the spectrum is
scale invariant for any case. This means that if the universe is
contracting, $n<1/3$ is required for the emergence of primordial
perturbation, while if the universe is in expansion, $n>1/3$ is
required, which corresponds that the early universe is dominated
by the component with $w<1$. Thus we reproduce the result of Ref.
\cite{M3}, however, by solving the motion equation (\ref{uk1}) of
perturbation mode on super sound horizon for any background
evolution of early universe. The effect of background evolution on
spectrum is reflected in the term $(\omega\eta)^{0.5-v}$ in
Eq.(\ref{uk}), which is just 1 for $z$ being exactly $3$.
In term of (\ref{uk}) and (\ref{v}), the dependence of tilt of
spectrum on $z$ and $n$ can be seen clearly.


The end of UV regime means the epoch at which the term
${\Phi\bigtriangleup^3\Phi\over a^6M^4}$ is replaced with
$\Phi\bigtriangleup\Phi$ in (\ref{I}), which means that the
universe is entering into the IR regime of HL gravity. This
requires ${k^2\over a^2M^2}\simeq 1$, thus $h_e\simeq M$, where
the subscript `$e$' denotes the end epoch of UV regime, since
$k^3=a^3hM^2$ for the perturbation mode just leaving the horizon.
This spectrum of $\Phi$ field can be inherited by that of
curvature perturbation in IR regime, which thus leads to the scale
invariant curvature perturbation. The efolding number for
primordial perturbation is defined as \be {\cal N} =
\ln\left({k_e\over k}\right),
\label{N}\ee where $k$ is the comoving wave number, which is equal
to the value at the time when the corresponding perturbation mode
leaves the sound horizon. This definition actually corresponds to
the ratio of the physical wavelength of perturbation mode
corresponding to the present observable scale to that at the end
epoch of UV regime. which is generally not equal to the efolding
number of scale factor. We substitute the comoving wave number
$k=ah^{1/3}M^{2/3}$ and $h\sim 1/a^{{1/ n}}$ into (\ref{N}), and
have ${\cal N}= (n-{1\over 3})\ln{({h\over h_e})}$, which is
consistent with the requirement of $n>1/3$ discussed. This result
indicates that, for fixed $n$, the resulting $\cal N$ depends on
the ratio $h$ to $h_e$, which must be large enough to match the
requirement of observable cosmology.

The efolding number $\cal N$ required is generally determined by
the evolution of standard cosmology after the UV regime ends.
In general, for simplicity, we assume that after the UV regime
ends the energy density of background field can rapidly
transferred into that of radiation, which will bring the universe
to an evolution of standard cosmology. We regard $M_e$ as the end
scale of UV regime, which approximately equals to the reheating
scale. In this case, the observation requires ${\cal N} \simeq
68.5+ \ln({M_e/M_P}) $ \cite{LL}, which is actually consistent
with that given by Ref. \cite{WMAP5}. It can be noticed that
$M_e\simeq \sqrt{MM_P}$, since $h_e\simeq M$.

\begin{figure}[t]
\begin{center}
\includegraphics[width=7.5cm]{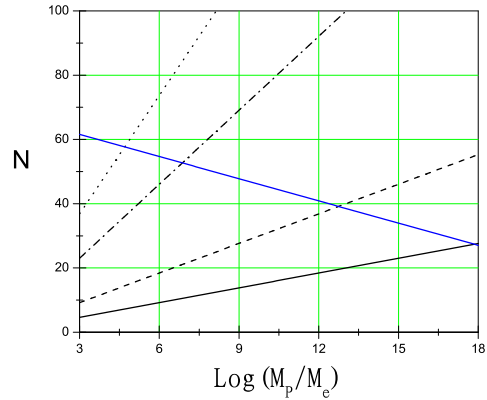}
\caption{ The figure of the $\cal N$ with respect to $\log
({M_P\over M_e})$. The black lines from lower to upper one
correspond to those for $n={2/3},\,1,\,2,\,3$, respectively. The
lower $M_e$ is at the time when the UV regime ends, the larger
$\cal N$ obtained is. The blue solid line is $\cal N$ required by
observable universe, which is determined by the evolution of
standard cosmology after the UV regime ends. The region above the
blue line is that with enough efolding number. }
\end{center}
\end{figure}


We plot the figure of the $\cal N$ with respect to $\log
({M_P\over M_e})$ in Fig.2, where that the UV regime begins at
$M_P$ has been set. We can see that for $1/3<n<1$, it seems
difficult to obtain the enough efolding number, unless the scale
$M_e$ at which the UV regime ends is quite low. For example, for
$n=2/3$, we have ${\cal N}\simeq 3$ for $M_e\sim 10^{15}$Gev,
while only when $M_e\sim 100$Gev, can the enough efolding number
be acquired. It is clear that if $n>1$, it is easier to have
enough efolding number suitable for observable universe, for
example, for $n=3$, we have $N\simeq 45$ for $M_e\sim 10^{15}$Gev,
while when $M_e\sim 10^{13}$Gev, $N>70$.

The period of $n>1$ corresponds to that of an accelerated
expansion, i.e. inflation. Thus in HL cosmology, though we can
obtain a scale invariant spectrum of primordial perturbation for
any expansion with $n>1/3$, but it seems that we still need a
period of inflation to obtain enough efolding number of primordial
perturbation. However, it is significant that, compared with
inflation with nearly exponential expansion, here $n\gg 1$ is not
required, for example, $n\simeq 3$ is enough, which helps to relax
the bounds for inflation model building. In principle, for
$1/3<n<1$, we can also consider some methods to obtain enough
efolding number, e.g. \cite{Piao0609}. In addition, it is also
interesting to explore above case in the bounce cosmology
\cite{B}.

In conclusion, it is showed by solving the motion equation of
perturbation mode for any background evolution of early universe
that the primordial perturbation can be generated naturally in UV
regime in HL cosmology for any expanding period of early universe
with $n>1/3$, which is scale invariant on large IR scale.
However, it seems that if we require the efolding number of
primordial perturbation suitable for observable universe,
$n\gtrsim 1$ still need to be satisfied, unless the scale of UV
regime is quite low. The motion equation of tensor perturbation in
UV regime is similar to that of scalar perturbation. Thus the
similar discussions can be applied.
This means
that in principle we can have a detailed compare of results with
recent observations \cite{WMAP5}, which will be considered. This
work might be interesting for motivating further studies for HL
cosmology.

\textbf{Acknowledgments} This work is supported in part by NSFC
under Grant No:10775180, in part by the Scientific Research Fund
of GUCAS(NO.055101BM03), in part by CAS under Grant No:
KJCX3-SYW-N2.

\end{document}